\documentclass[Afour,sageh,times]{sagej}

\usepackage{moreverb,url}
\usepackage{makecell}
\usepackage{multirow}

\usepackage[colorlinks,bookmarksopen,bookmarksnumbered,citecolor=red,urlcolor=red]{hyperref}

\newcommand\BibTeX{{\rmfamily B\kern-.05em \textsc{i\kern-.025em b}\kern-.08em
T\kern-.1667em\lower.7ex\hbox{E}\kern-.125emX}}

\def\volumeyear{2016}

\begin{document}

\runninghead{Roa Dabike et al.}

\title{The first Cadenza challenges: using machine learning competitions to improve music for listeners with a hearing loss}

\author{Gerardo Roa Dabike\affilnum{1},
Michael A. Akeroyd\affilnum{3},
Scott Bannister\affilnum{2},
Jon P. Barker\affilnum{4},
Trevor J. Cox\affilnum{1},
Bruno Fazenda\affilnum{1},
Jennifer Firth\affilnum{3},
Simone Graetzer\affilnum{1},
Alinka Greasley\affilnum{2},
Rebecca R. Vos\affilnum{1} and
William M. Whitmer\affilnum{3}}

\affiliation{\affilnum{1}University of Salford, UK\\
\affilnum{2}University of Leeds, UK\\
\affilnum{3}University of Nottingham, UK\\
\affilnum{4}University of Sheffield, UK}

\corrauth{Gerardo Roa Dabike
Acoustics Research Centre,
University of Salford,
Salford,
M5 4WT, UK.}

\email{g.roadabike@salford.ac.uk}

\maketitle

\section{Abstract}


It is well established that listening to music is an issue for those with hearing loss, and hearing aids are not a universal solution. How can machine learning be used to address this? This paper details the first application of the open challenge methodology to use machine learning to improve audio quality of music for those with hearing loss. The first challenge was a stand-alone competition (CAD1) and had 9 entrants. The second was an 2024 ICASSP grand challenge (ICASSP24) and attracted 17 entrants. The challenge tasks concerned demixing and remixing pop/rock music to allow a personalised rebalancing of the instruments in the mix, along with amplification to correct for raised hearing thresholds. The software baselines provided for entrants to build upon used two state-of-the-art demix algorithms: Hybrid Demucs and Open-Unmix. Evaluation of systems was done using the objective metric HAAQI, the Hearing-Aid Audio Quality Index. No entrants improved on the best baseline in CAD1 because there was insufficient room for improvement. Consequently, for ICASSP24 the scenario was made more difficult by using loudspeaker reproduction and specified gains to be applied before remixing. This also made the scenario more useful for listening through hearing aids. 9 entrants scored better than the the best ICASSP24 baseline. Most entrants used a refined version of Hybrid Demucs and NAL-R amplification. The highest scoring system combined the outputs of several demixing algorithms in an ensemble approach. These challenges are now open benchmarks for future research with the software and data being freely available.

\keywords{Music, Hearing Loss, Hearing Aids, Hearing Impairment, Machine Learning, Challenges}
\section{Introduction}
Most, if not all human cultures have music \citep{blacking1995music}. Music brings people together, shapes society and offers significant benefits to health and well-being \citep{macdonald2013music}. Hearing loss can detract from the listening experience, however. The World Health Organisation estimates that by 2050 2.5 billion people will have some form of hearing loss, with at least 700 million requiring treatment \citep{world2021world}. Hearing loss can lead to a range of challenges with music, including the inaudibility of quieter passages, poor or anomalous pitch perception, and difficulty in identifying and distinguishing lyrics and instruments \citep{hake2023development, moore2016effects, siedenburg2020can}. Therefore, it is essential to develop improved methods for processing music on hearing aids and consumer devices, enabling those with hearing loss to continue enjoying and benefiting from music.

The most common intervention for mild to moderately severe hearing loss is hearing aids. Many of these devices have music programs but efficacy is mixed \citep{Greasley2020, madsen2014music, looi2019music, vaisberg2019qualitative}. For example, \citeauthor{Greasley2020} (\citeyear{Greasley2020}) found that 68\% of users report difficulties when listening to music through their hearing aids. The issue is complicated because hearing aids are typically frequency-dependent, non-linear amplifiers to compensate for an individual's elevated thresholds, which must also allow for the rapid growth in loudness with low-intensity sound (loudness recruitment) and the potential discomfort from overcompensating louder sounds. These wide-dynamic range compression systems (WDRC) should make incoming sound audible and comfortable. WDRCs alter the temporal envelope of the signal, however, with the degree of change dependent on how quickly they react to dynamic fluctuations. They can also introduce dynamic artefacts such as a 'pumping' sensation. Hearing aids also have additional features such as speech enhancement, feedback management, wind-noise reduction and scene analysis. The settings of hearing aids, from the frequency-dependent gain to how quickly the compressor reacts to additional features, are predominantly optimised for speech,  and this means they may harm music, which has different spectral and temporal characteristics \citep{madsen2014music}.

Research into hearing aid processing and music perception has indicated some possible approaches to improve audio quality, although results are often mixed. \citeauthor{uys2012influence} (\citeyear{uys2012influence}) found that frequency compression, which shifts the spectrum and/or envelope of high-frequency information in the signal to more audible lower frequencies, improved the retrospective self-report of moderate to severe hearing-aid users. Although a later study found no statistically significant differences with frequency compression \citep{ahn2021influence}. \citeauthor{croghan2014music} (\citeyear{croghan2014music}) found that the quality of rock and classical music could be improved by using slow-acting rather than fast-acting WDRC. In contrast, \citeauthor{madsen2015effects} (\citeyear{madsen2015effects}) found  no significant overall effect of WDRC compression speed; there was no change in hearing-aid users' ability to hear individual instruments, although some participants found that slow-acting WDRC improved subjective clarity. These studies were based around more traditional signal processing approaches, however. Nowadays, machine learning is the dominant paradigm in new audio processing algorithms. While machine-learning techniques have shown improvements in speech intelligibility for hearing-aid algorithms e.g. \citep{bramslow2018improving, akeroyd20232nd}, there is a gap in knowledge about how machine learning can improve perceived audio quality of music for those with a hearing loss. 

Sound engineering approaches such as remixing, have potential to improve audio quality for those with a hearing loss. \citeauthor{benjamin2023exploring} (\citeyear{benjamin2023exploring}) explored how listener preference was changed by altering aspects of pop music such as lead-to-accompaniment level ratio, low-to-high frequency spectral energy balance and transformed equalisation. Elevated lead-to-accompaniment level ratio and music that was spectrally sparser was preferred by those with hearing loss. 

In signal processing, many significant advances have been driven by open machine-learning challenges (competitions) e.g. \citep{barker2017, liberman2020human, fabbro2024sound}. The required features of a rigorous, well-designed challenge are: 

\begin{itemize}
    \item A common task with clear rules to constrain the solution space, ensure a fair competition and address a defined research question.
    \item Software tools including a baseline that gives a solution to the problem for entrants to build upon.
    \item Three common datasets (training, validation, and evaluation). Challenge entrants use the training data to build algorithms that replace parts of the baseline. After submission, every entry is tested and ranked using the independent evaluation data, which is often provided just before submission deadline.
    \item Rules for (1) what datasets and pre-trained models can be used in training and development, and (2) restrictions on what can be modified or allowed, are set to allow a fair competition.
    \item A strict submission deadline, typically 4-6 months after launch.
    \item A workshop or special conference session to announce results and bring competitors together to share knowledge and shape future challenges.
\end{itemize}
By providing a challenge infrastructure, including open databases for machine learning and specialized software tools, challenge organizers can significantly lower barriers that may have historically prevented out-of-field researchers from engaging in a topic. Challenges have also been shown to foster collaboration across disciplines, attracting a wider and more diverse range of researchers who contribute novel approaches to the field. Challenges also create an important legacy in the form of open benchmarks for future research. For the Cadenza project, both challenges were free to enter, with all the materials being provided at no cost to encourage as many entrants as possible.

In this paper we describe the Cadenza project: the first application of the challenge methodology to the problem of improving audio quality of music for listeners with a hearing loss. Two challenges are reported, the primary difference being that the first Challenge (CAD1) \citep{CAD1} from 2023 was for listening over headphones, and the second (ICASSP24), \citep{ICASSP2024}), was for listening over loudspeakers. The tasks targeted demixing of stereo music signals followed by remixing, because such as system could help with known problems for music listening and hearing loss \citep{Greasley2020}. For example, an amplification of the vocals between demix and remix could help with lyric intelligibility. Demixing was also chosen because there was an existing research community to tap into from previous challenges -- the  Signal Separation Evaluation Campaigns (SiSEC) 2015-18 \citep{Ono2015, liutkus20172016, stoter20182018} and the Music Demixing Challenges MDX2021 and SDX2023 \citep{mitsufuji2022music, fabbro2024sound} –- though none of those considered listeners with hearing loss. Building on these previous challenges, the demixing was into vocal, drums, bass and other instrument stems (VDBO). In both challenges the measure of success was to score as high as possible on the Hearing Aid Audio Quality Index (HAAQI) \citep{kates2015hearing}.

To be successful in addressing the research questions, challenges require rigour and care on the part of the organizers. There is a difficult balance to strike between making the challenge too easy, which restricts innovation, or too hard, which can dissuade teams from entering. The considerable materials supplied need to be sufficient, the rules and scoring fair. Difficult decisions have to be made before launch when there is considerable uncertainty about what teams might do. For this reason, the paper gives in-depth details of the materials and methods developed for the two challenges, outlining the reasoning behind the scenarios, rules, baselines and data. This is followed by analysis of the entries and objective evaluations of their success. The paper finishes with a critique of the challenges and how this is informing future work.
\section{Materials and Methods} \label{evaluation-materials}

\subsection{Overview}

\begin{figure}
    \centering
    \includegraphics[width=\linewidth]{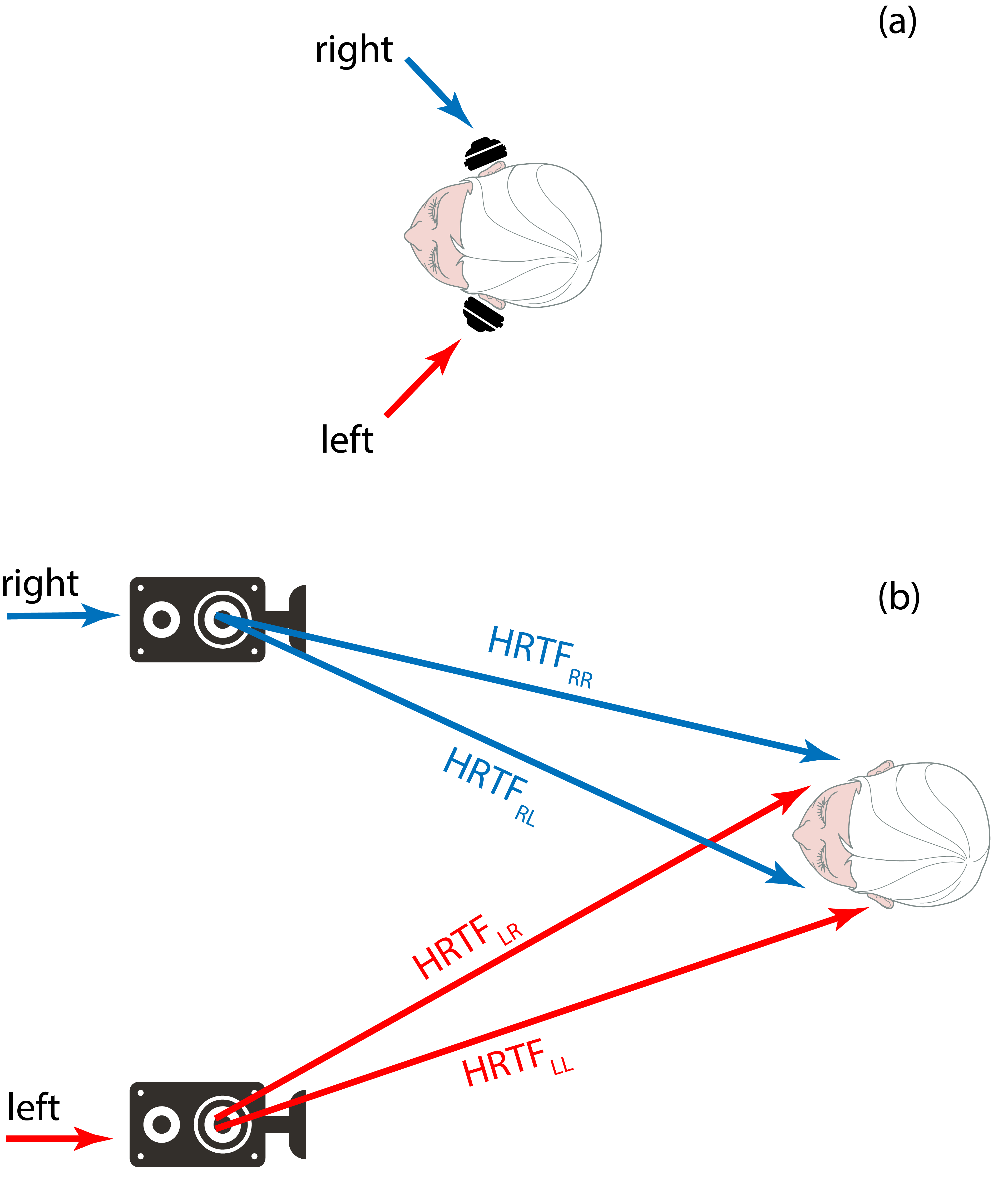}
    \caption{The scenarios for (a) CAD1 headphone listening and (b) ICASSP24 loudspeaker listening via hearing aids. HRTF, Head-Related Transfer Function.}
    \label{fig:crosstalk}
\end{figure}

The two scenarios were based around listening to music over (1)  headphones without hearing aids, and (2) stereo loudspeakers using hearing aids - see Figure \ref{fig:crosstalk}. For CAD1, the signals to be processed were the left and right signals being fed to the headphones. For ICASSP24 the left and right signals were from the hearing aid microphones at each ear. In both challenges entrants were challenged to create a system that could rebalance the levels of the vocal, drums, bass and other instruments (VDBO). This would then allow personalised mixes for people with a hearing loss. The VDBO representation was chosen because of its use in previous demix research.

\begin{figure}
    \centering
    \includegraphics[width=\linewidth]{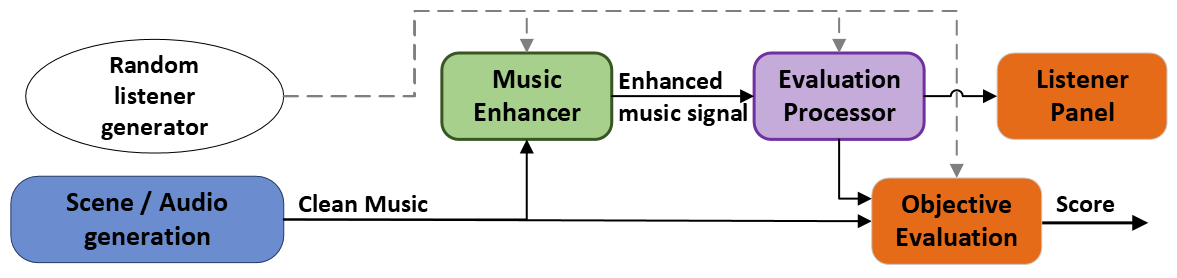}
    \caption{General structure of the challenges}
    \label{fig:Baseline}
\end{figure}

Figure \ref{fig:Baseline} shows the general structure of the challenge. Entrants were presented with scenes (see blue box) containing a music extract to process and metadata giving the rendering requirements for the sample. (For example, in ICASSP24 the metadata specified gains for the isolated sources in the mixture.) Additionally, a random listener (white oval) was selected, giving a pair of audiograms to allow personalisation of the signal processing and evaluation. A challenge rule specified that entrants were only allowed to modify the Music Enhancer (green box). The Evaluation Processor (lilac box) prepares the samples for either objective evaluation using HAAQI or perceptual testing via the listening panel.

Table \ref{T1} compares the differences between the CAD1 and ICASSP24 challenges. In CAD1, the music to be processed was the stereo signals being fed to a pair of headphones. In contrast for ICASSP24, the music came from left and right hearing aid signals when listening over a stereo loudspeaker pair. This meant that for ICASSP24, the music to be processed was a mixture of both the right and left loudspeaker signals -- see Figure \ref{fig:crosstalk}. The sound propagation from the loudspeakers to the hearing aid microphones were modelled using Head-Related Transfer Functions (HRTFs). How the the left and right signals from the loudspeakers combine at the left and right ears is dependent on wave diffraction, reflection and interference around the shoulders, head, ears and hearing aids. At some frequencies and azimuths they may add, at others they may subtract. Consequently, the strength of the left and right VDBO components at the ear are different compared to the original stereo tracks, posing additional complexities for ICASSP24 systems compared to CAD1 and previous demixing challenges.

\begin{table}[!h]
\small\sf\centering
\caption{Differences between CAD1 and ICASSP24 challenges\label{T1}}
\begin{tabular}{lll}
    \toprule
        &  \textbf{CAD1}  &    \textbf{ICASSP24} \\
    \midrule
        \multirow{2}{2.3cm}{Listening via} & \multirow{2}{2.3cm}{Headphones} & \multirow{2}{2.3cm}{Loudspeakers} \\
        && \\

        \multirow{2}{2.3cm}{Gains applied\\before remixing} & \multirow{2}{2.3cm}{No} & \multirow{2}{2.3cm}{Yes}  \\
        &&\\
        
        \multirow{2}{2.3cm}{Evaluation} & \multirow{2}{2.3cm}{HAAQI and\\listening panel} & \multirow{2}{2.3cm}{HAAQI}\\
        &&\\
    \bottomrule
\end{tabular}
\end{table}

For CAD1, there was both objective and perceptual evaluation, whereas for ICASSP24 only objective evaluation was done. ICASSP24 did not include listening tests because ICASSP grand challenges run on a short timescale, leaving no time to carry out experiments. In this paper, only the objective evaluation for the two challenges are given. The CAD1 listening panel experimental design and results will be presented in a subsequent paper.

A final difference between the two challenges, was that in CAD1 there were no gains applied to the VDBO signals before remixing back to stereo, whereas there was in ICASSP24. The gains were added to make the ICASSP24 more challenging because teams struggled to beat the baselines in CAD1. Changing the levels between the VDBO components should make artefacts created in the processing less likely to be masked. Furthermore, it will highlight cases where separation is imperfect. For example, in CAD1 if some of the drums was wrongly put into the bass track, then when the VDBO were summed together to give the stereo remix, the demix failure would be hidden. In ICASSP24, when there were different gains for the drums and bass track, this demix failure would result in the stereo remix being wrong. 

\subsection{Overview of databases}
Fundamental to machine learning using Deep Neural Networks (DNNs) is having access to very large amounts of data. The datasets were split into training, validation and evaluation sets. Both the training and validation datasets were provided when the challenge launched, and were used by the teams to develop the signal processing systems. 

Training data is used to update the machine learning algorithm whereas validation data is used to monitor the progress of the training and check for issues such as overfitting. The evaluation set tested generalisation of systems to different music and listeners. The evaluation data was made available only a few weeks before the submission deadline. The challenge rules stated that teams should not use the evaluation data to improve their system, but they should simply pass the signals through their systems and submit these for evaluation. The evaluation data only contained the stereo mixed music tracks. 

With machine learning, having access to large private datasets can give teams an unfair advantage. For this reason, the challenge rules specified that teams could only use datasets and pre-trained models that were in the public domain. However, entrants were allowed to augment the data using simple processing. For example, they could randomize the stems, flip the right and left channels, apply SpecAugmentation \citep{park2019specaugment} and pitch shifting. Such augmentation is very common in machine learning to create more robust systems.

\subsection{Listener databases}

Each music extract needed to be personalised to allow for the hearing acuity of a target listener. The hearing was characterised by bilateral pure-tone audiograms at the standardised frequencies of 250, 500, 1000, 2000, 3000, 4000, 6000, 8000\,Hz. The bandwidth of music is wider than this, but we were limited by available datasets of audiograms. The datasets were measured, anonymous audiograms from bilateral hearing aid users. Hearing loss levels at each frequency were limited to 80\,dB to be consistent with the training dataset from the Clarity Project \citep{CEC2}. This limit was applied in Clarity because (i) the hearing loss model they used produces unrealistic signals for large impairments and (ii) the headsets they used in listening tests could not reproduce high enough levels to compensate for large impairments. 

Hearing loss severities were based on the mean, better-ear 4-frequency (500, 1000, 2000 and 4000\,Hz) hearing loss criteria \citep{stevens2013global}. These were no impairment (0-19\,dB), mild (20-34\,dB), moderate (35-49\,dB), moderately severe (50-64\,dB), severe (65-79\,dB) and profound impairment ($>=$ 80\,dB). 

The datasets were as follows:

\begin{itemize}
    \item Training: 83 audiograms from the Clarity Project \citep{CEC2}. These correspond to real but anonymised audiograms drawn from the participant database of the Scottish Section of Hearing Sciences at the University of Nottingham. By our better-ear hearing loss categorisation there were no people with no impairment, 17 people with mild, 44 with moderate, 22 with moderately severe and none with severe. 
    \item Validation: 50 audiograms drawn from the dataset by von Gablenz et al. \citep{Von2017}. The 50 audiograms were randomly selected to have the same distribution as the training set. First, audiograms were filtered using better-ear 4-frequency hearing loss criteria, with thresholds between 20 and 75 dB. Then, the audiograms were randomly chosen to maintain the same distribution per band as in the training set. This set had an equal male-female distribution. The distribution by our categorisation was no people with no impairment, 24 with mild, 22 with moderate, 4 with moderately severe and 0 with severe.
    \item Evaluation: 53 audiograms; 52 listeners with a hearing loss were recruited for the Cadenza listening panel by the University of Leeds. An additional audiogram for no impairment (0 dB loss at all frequencies) was also included to evaluate systems for no impairment conditions. The distribution by our categorisation was as 3 listeners with no impairment, 13 with mild, 17 with moderate, 19 with moderately severe and 1 with severe.
\end{itemize}

\subsection{Music datasets}

There are established publicly-available datasets that have become the benchmark data for demixing challenges. These were therefore used for CAD1 and ICASSP24 for comparability with previous work. The music for training, validation and evaluation used the standard splits for MUSDB18-HQ \citep{musdb18-hq}, giving 86, 14 and 50 stereo tracks respectively. MUSDB18-HQ contains isolated stems for vocals, drums, bass and other (VDBO), as well as stereo mixes. The music is mostly Western pop/rock with a small amount of reggae, rap, heavy metal and electronic.

An independent validation set was constructed by randomly selecting 50 tracks from the MoisesDB dataset \citep{moisesdb}, while maintaining the same genre distribution as the evaluation split of MUSDB18-HQ. This new validation set was included because many pretrained models that use MUSDB18-HQ, incorporate the MUSDB18-HQ validation split as part of their training.

\begin{figure}
    \centering
    \includegraphics[width=\linewidth]{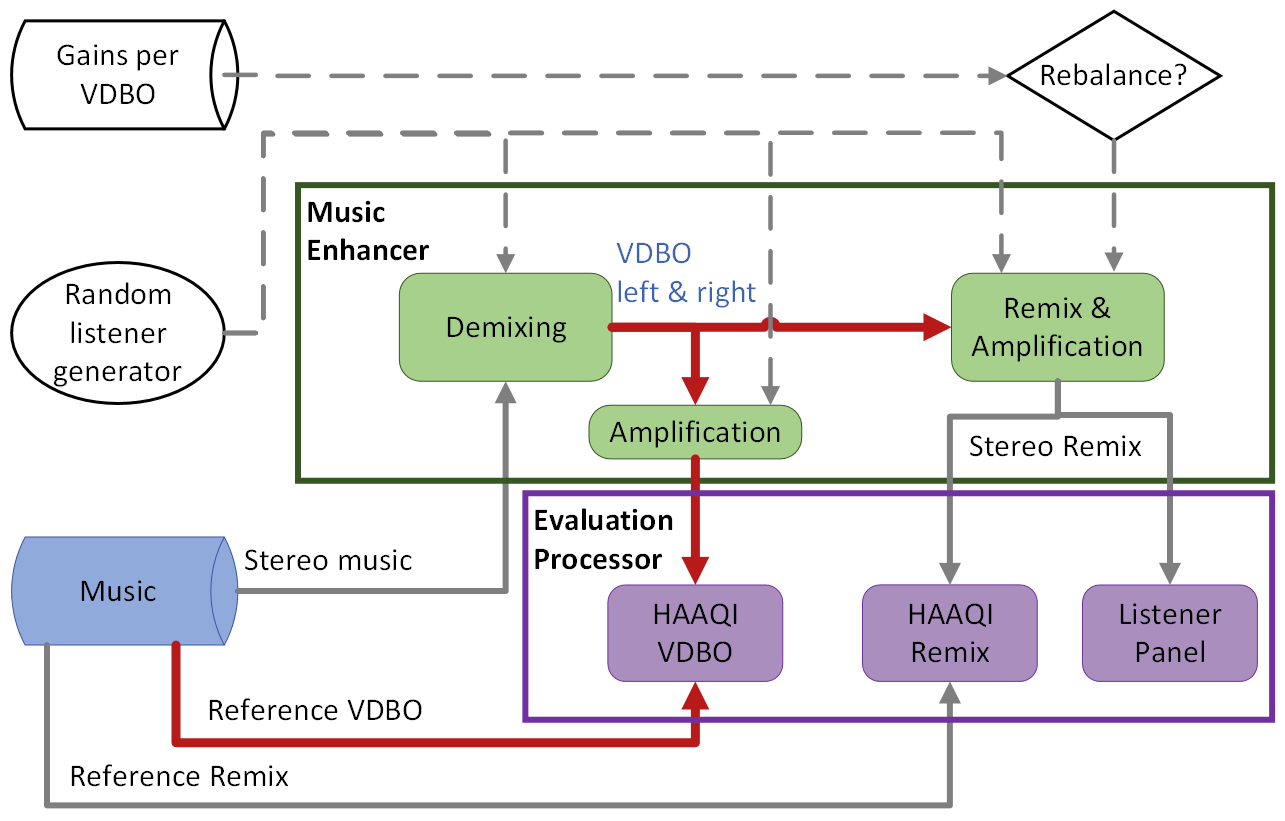}
    \caption{Baseline Architecture for CAD1 and ICASSP24.}
    \label{fig:general_diagram}
\end{figure}


\subsection{Baseline software tools}

The baseline is a complete software system that can run the task. It includes a solution to the problem in the music enhancer for entrants to beat. Figure \ref{fig:general_diagram} shows the architecture. The problem was presented as a demix/remix task, with a view to allowing listeners to rebalance and personalise a mix. Systems needed to take stereo pop/rock music and demix it into VBDO signals. Gains could then be applied to these four signals before they were remixed back to stereo. This is similar to previous demix challenges \citep{stoter20182018, mitsufuji2022music, fabbro2024sound}, except entrants could allow for a listener's hearing loss in the processing. An additional novelty compared to previous demix challenges was that the evaluation metric is one tailored for listeners with hearing loss and hearing aids. HAAQI, the Hearing-Aid Audio Quality Index \citep{kates2015hearing}, was used. Its the only published computational-amenable metric for audio quality of music for listeners with a hearing loss listening through hearing aids.

While the baseline did demixing, the challenge rules did allowed entrants to submit stereo audio from an end-to-end system without an explicit demixing stage. However, no entrant chose to do this.

Two baseline demix algorithms were given to entrants. These were out-of-the-box pretrained audio source separation algorithms with no retraining. One used the Hybrid Demucs model \citep{defossez2021hybrid} (HDemucs), which employs a U-Net architecture to combine both time-domain and spectrogram-based audio source separation. The other used the Open-Unmix model \citep{stoter19}, which just uses spectrograms. HDemucs is probably the most powerful demixing algorithm in use, whereas Open-Unmix is simpler to implement and train.

The music enhancer also needed a frequency-dependent amplification stage to correct for the raised auditory thresholds due to hearing loss. NAL-R \citep{byrne1986national} was used to match the default amplification applied to the reference signal during the HAAQI evaluation (see below). As HAAQI compares the processed signal to a reference, the frequency-dependent amplification stage needs to be the same to maximise HAAQI scores.

\subsection{Latency and model size}
For a signal processing system to be used with live music on a hearing aid, it needs to operate with low latency. This restricts the machine learning approaches that teams can take. However, listening to recorded music is also very common, so there are scenarios where latency is not a concern. For these reasons, the rules allowed both causal and non-causal systems. For causal processing, the challenge rules restricted systems to only use input samples less than 5 ms in the future, which is about the tolerable delay between a hearing aid's output and the original acoustic signal \citep{vatakis2006audiovisual}.

Hearing aids can not currently host the huge deep neural networks that are common in audio processing. However, the challenges did not place a limit on the model size or computing resources being used by systems. The reason for this is that a common innovation route in machine learning is to first produce solutions that are computationally expensive and then apply methods such as knowledge distillation \cite{gou2021knowledge} to reduce the resources required while maintaining performance.

\subsection{Rendering and remix metadata for ICASSP24}

For the ICASSP24 challenge, the scene generator had to simulate loudspeaker reproduction. This was done by applying HRTFs (Head Related Transfer Functions) measured in anechoic conditions - see Figure \ref{fig:crosstalk}. The scene generator randomly selected one of the 16 subjects from the OlHeaD-HRTF dataset \citep{DenkHRTF}, the azimuth angle of the listener's head, and then extracted the appropriate HRTFs from the OlHeaD-HRTF dataset. Modern hearing aids typically have multiple microphones to allow for some beam forming, but here just the front left and right microphones of the hearing aids were used.

Listeners were modelled to have a variety of head orientations around the azimuth range for standard stereo loudspeaker reproduction (i.e. around \textpm30\textdegree). This was to simulate non-perfect stereo reproduction. Angles were all nine combinations from \textpm22.5\textdegree, \textpm30\textdegree and \textpm37.5\textdegree.

Each music track was divided into several consecutive 10-second excerpts, ensuring that no silent segments were selected. Then a HRTF pair was applied to each excerpt. This means that two excerpts from the same track will have different pairs of HRTFs applied, thus requiring separation models to be robust under varying HRTF conditions and for different songs.

For ICASSP24, the generator also randomly set the gains to be applied to each VDBO stem before the remix to stereo. Unfortunately, there was little prior knowledge to guide what might be the preferred gains for the VDBO stems, and furthermore these would vary with listener preference and the music. Consequently, random gains were used to bracket significant changes in the stereo remix to create systems that could enable any remix a listener might ask for. First the number of VDBO stems that had their gain altered was randomly chosen using a uniform probability distribution (i.e. 1, 2 or 3). Then the gains were chosen for each of these tracks from \textpm10\,dB, \textpm6\,dB and \textpm3\,dB. Again a uniform probability random process was used.

\subsection{Evaluation}

Objective evaluation was done using HAAQI \citep{kates2015hearing}. For the VDBO evaluation, HAAQI was calculated for 8 stems (V, D, B and O left, and V, D, B and O right) and then an average taken. For evaluating the remix stereo, HAAQI was calculated for the left and right signals, and an average of the two HAAQI scores used.

HAAQI was developed as a perceptual model and so is relatively slow to compute and non-differentiable. These make it harder to incorporate into machine learning efficiently. Furthermore, it is an intrusive metric (double-ended) and therefore requires a reference signal. This reference needs a frequency-dependent amplification to correct for raised hearing thresholds. Whatever amplification scheme is chosen, this must then also be replicated in the music enhancer to maximise HAAQI scores. In both  CAD1 and ICASSP24, a linear FIR filter based on a NAL-R prescription was used \citep{byrne2001nal}. This used a public-domain implementation that was available. While most hearing aids might use a bank of dynamic range compressors each operating over different bandwidths, the best settings for these compressors is disputed (see Introduction).

For CAD1, 49 out of the 50 music tracks in the MUSDB18-HQ evaluation split were used. Because of the subjective evaluation by listeners, one track was excluded due to offensive words in the lyrics. To keep the submission size within reasonable bounds (around 23 GB; 4 VDBOs for the left channel, 4 VDBOs for the right channel, and 1 remix for each listener), entrants were required to submit 30-second extracts. These extracts were selected randomly, ensuring that all VDBO stems were active at some point. Each extract was processed for all 53 listeners obtaining N = 2,597 processed extracts per system. For ICASSP24, all 50 evaluation tracks from MUSDB18-HQ were used. The tracks were segmented into consecutive 10-second extracts, resulting in 960 audio segments. For these 960 extracts, the music was paired with a random HRTF and a random gain. To keep the submission package around 20 GB, each of these were processed for 20 random listeners from the pool of 53 listeners, giving N = 19,200 audio examples tested per system.  
\section{Results}

\subsection{CAD1 Challenge}

Seven entries, two baselines and a do-nothing system were evaluated. Table \ref{tab:CAD1} summarises the different approaches of the systems for CAD1 and the average HAAQI scores.

\begin{table*}[!t]
    \small\sf\centering
    \caption{Overview of system approaches and scores for CAD1 challenge. Total 10 systems with 2,597 processed audios per system. ``Original'' remix means using the gains specified for the original music. * indicates a refined version of the algorithm used. HAAQI scores are averaged over the evaluation set with $\pm$ standard deviations, with VDBO being for the separated VDBO stems and remix for the output stereo. Finally, average HAAQI results across the evaluation set are shown with standard deviations as \textpm~values (N=2,597). Values in bold show the highest scores. E21 is a do nothing system where the processed signals are equal to the original signals with no amplification.}
    \begin{tabular}{ l c c c c c }
    \toprule
           
     \textbf{System}	& \textbf{Separation} & \textbf{Remix} & \textbf{Amplification} & \textbf{HAAQI VDBO} & \textbf{HAAQI remix}\\
    \midrule
        E01, Baseline 1 & HDemucs & Original & NAL-R & \textbf{0.255 $\pm$ 0.041} & \textbf{0.706 $\pm$ 0.196} \\
        E02, Baseline 2& OpenUnmix & Original & NAL-R & 0.225 $\pm$ 0.029 & 0.638 $\pm$ 0.161 \\
    \midrule
        E05 & OpenUnmix* & Original & NAL-R & 0.094 $\pm$ 0.014 & 0.677 $\pm$ 0.186 \\
        E12 & HDemucs & Rebalanced & Multiband compressor & \textbf{0.255 $\pm$ 0.041} & 0.684 $\pm$ 0.205 \\
        E14 & HDemucs & Original & NAL-R* & 0.203 $\pm$ 0.029 & 0.530 $\pm$ 0.226 \\
        E15 & OpenUnmix & Original & NAL-R* & 0.183 $\pm$ 0.023 & 0.475 $\pm$ 0.191 \\
        E21 & - & Original & None & 0.421 $\pm$ 0.216 & 0.440 $\pm$ 0.234 \\
        E16 & Spleeter \citep{hennequin2020spleeter} & Original & NAL-R* & 0.135 $\pm$ 0.027 & 0.270 $\pm$ 0.148 \\
        E17 & HDemucs & Mid-Side EQ & NAL-R + compressor & 0.236 $\pm$ 0.033 & 0.276 $\pm$ 0.105 \\
        E22  & HDemucs & Rebalanced & NAL-R & 0.195 $\pm$ 0.039 & 0.217 $\pm$ 0.109 \\
    \bottomrule
    \end{tabular}
    \label{tab:CAD1}
\end{table*}

Eight systems used either the HDemucs or OpenUnmix models for source separation. One of those (E05) refined the OpenUnmix by using a sliced Constant-Q Transform (sliCQT) \citep{holighaus2012framework} with the Bark scale, a neural network architecture that used a convolutional denoising autoencoder (CDAE) \citep{holighaus2012framework, grais2021multi}; and all targets were trained together with combined loss functions like CrossNet-Open-Unmix (X-UMX) \citep{sawata2021all}.

Six systems did not alter the remix from the baseline, meaning the demixed VDBO stems were simply added together to get the stereo output. Those that did alter the balance between the VDBO stems included E17. This system applied a mid-side equalisation, with a new left signal L' and new right signal R' given by:
\begin{equation}
\begin{array}{lr}
L' = G(M) + H(S) \\
R' = G(M) - H(S)
\end{array}
\end{equation}
Where M is the mid signal and S the side signal given by:
\begin{equation}
\begin{array}{lr}
M = (L+R)/2 \\
S = (L-R)/2
\end{array}
\end{equation}
The function \textit{G()} was two parallel filters that reduced components in the mid below 2\,kHz by 2\,dB to attenuate frequencies that were not part of the lead vocals. The function \textit{H()} was three parallel filters to increase the components in the side between 2\,kHz and 6\,kHz by 3\,dB to help with binaural unmasking.

Another two systems that changed the balance in the remix were E22 and E12. Both used methods to increase the prominence of the vocal track. For example E22 used gains of +7.6, -8.0, -4.4, and -4.4\,dB for the VDBO stems when all were not silent. E12 decreased the level of the non-vocal tracks for people with moderate or severe hearing loss.

Because the objective metric HAAQI is intrusive, meaning it compares the processed signal to a reference, any system that changed the EQ or balance of the VDBO stems before remix would decrease the HAAQI scores. It is assumed entrants did this to increase the scores in the listening tests.

Four systems used the NAL-R amplification provided in the baseline. The exceptions were: (1) E12, which used a multiband compressor. (2) E14 and E15, which used a linear filter like NAL-R but decreased the low-frequency attenuation in original NAL-R algorithm. The 250 and 500\,Hz bands were increased by 16 and 7\,dB respectively. The intention here was to increase the bass, but this had the unintentional consequence of limiting the high frequency amplification, which is where hearing loss is usually most significant. This happens because signals were peak normalised in the time domain across frequency to prevent clipping. (3) E16 applied a Butterworth bandpass filter with -3\,dB points at 250\,Hz and 18.5\,kHz. As the HAAQI reference used NAL-R, any departure from this amplification would naturally decrease the objective evaluation, but could have improved listening test scores.

The HAAQI scores averaged for the VDBO stems in Table \ref{tab:CAD1} need to be read with some caution, because applying HAAQI to individual stems is untested and the metric was designed for complete music. The objective scores for VDBO show that Baseline 1, which used HDemucs, scored higher than other systems. In setting up the challenge, it was hypothesised that the increased hearing thresholds that occur with hearing loss might have been allowed for in the source separation algorithms and therefore lead to improved performance. For instance, artefacts created during source separation might fall below the hearing threshold. But no system exploited this possibility. There have been many demixing challenges, see Introduction, which meant the state-of-the-art approaches used in the baseline were hard to beat.  

\begin{figure}
    \centering
    \includegraphics[width=\linewidth]{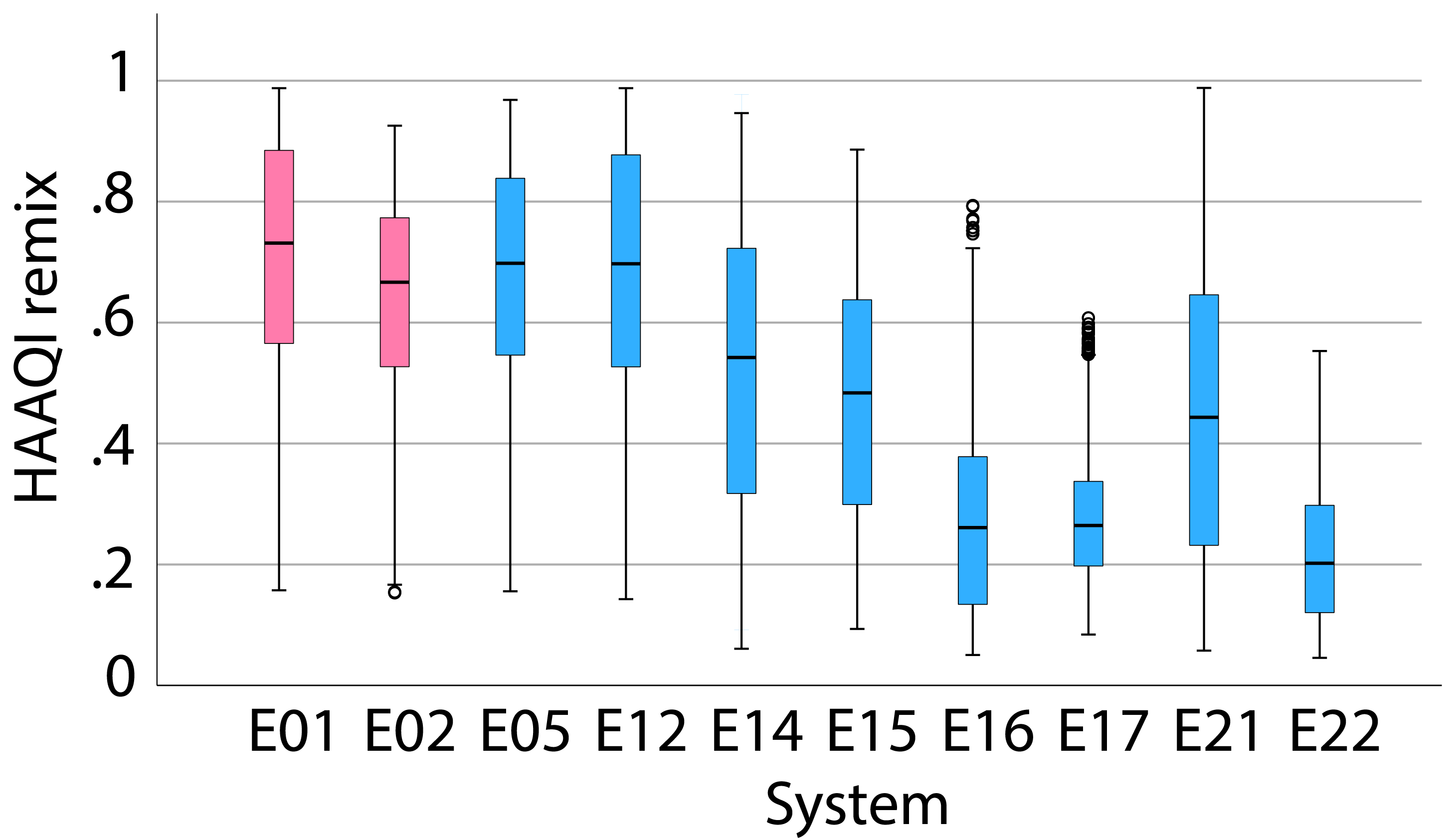}
    \caption{HAAQI scores for remix (downmix) vs system for CAD1. Baseline systems shown in pink.}
    \label{fig:CAD1boxplot}
\end{figure}

The objective scores for the remix stereo are shown in Table \ref{tab:CAD1} and a box-plot shown in Figure \ref{fig:CAD1boxplot}. The data did not meet the assumptions needed to use an ANOVA. For example, the dependent samples were not drawn from a normally distributed population with evidence of a ceiling effect where HAAQI=1 for some teams. Consequently, the following is an analysis of main effects using non-parametric approaches.

A one-way Kruskal-Wallis test with HAAQI values as the dependent variable and the systems as the independent variable was used to test whether differences between the scores for the ten systems were significant. This was indeed the case (N = 25,970, df=9, \textit{H}=12,824, p$<$0.001, $\eta ^2$=0.49) with a very large effect size. Pairwise comparisons show that most systems are significantly different from each other (p$<$0.001 for pairs with significant difference, except E05-E01 where p=0.02. Bonferroni correction for multiple tests applied). The three pairs of systems with no significance difference were: E16 and E17 (p=1); E05 and E12 (p=1); and E12 and E01 (p=0.1).

Whether system performance varied with hearing loss severity was tested. The listener audiograms were coded into a 5-value ordinal variable: no impairment, mild, moderate, moderately severe and severe. In this case the average of the left and right audiogram was used rather than the better ear to evaluate the severity. This was done because the HAAQI scores were an average of the HAAQI values for the left and right signals. The Spearman’s rho between HAAQI and hearing loss severity was -0.540 (N=25,970; p$<$0.001). This means the HAAQI scores were lower for those with worse hearing loss, a trend seen for all systems. This trend explains 29\% of the rank variance.

Overall, for the remix HAAQI scores, the HDemucs Baseline E01 system had a higher score than the other systems. As noted above, the 5 systems that applied different remix or amplification systems were almost bound to score lower on HAAQI. These entrants used different amplification approaches to improve scores in the listening panel evaluation.

The lack of HAAQI improvement over the baseline from CAD1 entrants indicated a need to run a challenge that offered more chance of bettering the baseline. This led to the ICASSP24 challenge. Specifically, the use of loudspeaker reproduction in ICASSP24 meant that out-of-the-box demix algorithms would perform worse because of the frequency-dependent mixing of the left and right music signals and gains would highlight bleed between demixed components. This also motivated a push for causal systems because non-causal approaches would not work on hearing aids.

\subsection{ICASSP24 Challenge}

There were 17 systems entered from 11 teams. Table \ref{tab:results_icassp} summarises the average HAAQI scores for the different systems. Nearly all differences between the system scores in the table were statistically significant, but some had very small effect sizes (see later for statistical analysis). The table also summarises the approaches for the different systems. Nine systems beat the best baseline, and the discussions below of the non-causal systems will concentrate on these.

\begin{table*}[!t]
    \small\sf\centering
    \caption{Overview of the approaches for ICASSP challenge. A: system with data augmentation. S: system with supplementary data. B: second submission. * indicates a refined version of the algorithm was used. ? indicates amplification model not specified in technical report and so assumed to be NAL-R from the baseline. Finally, average HAAQI results across the evaluation set. Standard deviation shown as \textpm~value, N=19,200. Values in bold show systems that performed better than T01 Baseline.}

    \begin{tabular}[t]{l c c c}
    \toprule
         \textbf{System}	& \textbf{Separation} &
         \textbf{Amplification} & \textbf{HAAQI} \\
    \midrule
    \textbf{Non-Causal}  & & &\\
    \midrule
        T01, Baseline & HDemucs & NAL-R & 0.570 $\pm$ 0.185 \\
        T02, Baseline & OpenUnmix & NAL-R & 0.511 $\pm$ 0.153 \\
    \midrule
         T47 \citep{T022} & Ensemble & NAL-R & \textbf{0.632 $\pm$ 0.177} \\
         T22 & Ensemble & NAL-R & \textbf{0.631 $\pm$ 0.173} \\
         T03-S \citep{T003} & HDemucs* & NAL-R & \textbf{0.593 $\pm$ 0.186} \\
         T03 \citep{T003} & HDemucs* & NAL-R & \textbf{0.592 $\pm$ 0.185} \\
         T11-A \citep{T011} & HDemucs* & NAL-R & \textbf{0.586 $\pm$ 0.188} \\
         T18 \citep{T018} & U-Nets + DPRNN& ? & \textbf{0.585 $\pm$ 0.183} \\
         T11 \citep{T011} & HDemucs* & NAL-R & \textbf{0.580 $\pm$ 0.186} \\
         T12 & HT-Demucs \citep{rouard2023hybrid} & ? & \textbf{0.573 $\pm$ 0.182} \\
         T46 \citep{T046} & HDemucs* & NAL-R & \textbf{0.570 $\pm$ 0.185} \\
         T25 & HDemucs & Compressor + NAL-R* & 0.561 $\pm$ 0.163 \\
         T31-A  & HT-Demucs* & ? & 0.543 $\pm$ 0.169\\
         T42 & HDemucs & ? & 0.543 $\pm$ 0.173\\
         T42-A & HDemucs & ? & 0.534 $\pm$ 0.172\\
         T31  & HT-Demucs* & ? & 0.530 $\pm$ 0.174\\   
         T09-B & HDemucs & ? & 0.479 $\pm$ 0.132\\         
         T09 & HDemucs & ? & 0.478 $\pm$ 0.135 \\
    \midrule
    \textbf{Causal} &\\
    \midrule
         T16 & k-means & ? & 0.144 $\pm$ 0.018\\
    \bottomrule
    \end{tabular}     
    \label{tab:results_icassp}
\end{table*}

The baselines were trained on the original stereo music and not on the hearing aid signals. Consequently, it was anticipated that retraining an established source separation system on the hearing aid signals would be sufficient to improve scores. Examples of teams doing this were T11 and T46.

The highest scoring system T47 took an ensemble approach, with the output of the separation algorithm being an average of three systems. These were pretrained versions of Dual-Path TFC-TDF UNet \citep{chen2024music}, HDemucs, and a version of MDX-Net \citep{kim2021kuielab} only trained on the MUSDB18-HQ dataset. These were then fine tuned on the ICASSP24 dataset. T22 took a similar ensemble approach but one of the two pretrained model used the label noise dataset from the Sound Demixing Challenge 2023 \citep{fabbro2024sound}, which was outside the rules of the ICASSP24 challenge.

There were some refinements within systems that built on established architectures. T03 and the version trained on supplementary data (T03-S) added 15\% of the original stereo into the final mix using a skip connection that bypassed the demix/remix. The intention was to restore components that get lost in the demix/remix process. T11 introduced a modified logit function intended to create a larger gradient for hard-to-learn examples. This used self-knowledge distillation with progressive refinement of target (PS-KD) \citep{kim2021self}. An ablation study showed this modified logit produced a very small improvement in HAAQI of 0.009. T46 replaced the original complex ratio mask in HDemucs with a deep filter \citep{T046}.

There were also some refinements on how the training data was used to improve learning. Augmentation achieved a very small improvement in HAAQI of 0.006 for T11-A vs T11. This team also explored curriculum learning where initial training was on easier examples in the training set before moving onto the harder case after a set number of epochs. This produced only a very small improvement in HAAQI of 0.002, however.

Only T25 attempted to improve the amplification stage of the processing. As the reference signal in the objective evaluation involved amplification using NAL-R, this could only reduce the HAAQI scores. However, it is worth noting that NAL-R cannot account for all the non-linear level-dependencies typical with hearing loss, such as loudness recruitment, and so the approach of T25 might result in improved scores in listening tests, but that remains untested.

The causal system T16 did not score as well as the non-causal approaches. Previous demixing research has mostly focused on non-causal approaches, so there were preexisting, more refined non-causal approaches to build upon for the Cadenza challenge. Furthermore, it would be expected that a causal method would score lower because any machine learning algorithm has less input information to work from compared to non-causal techniques. T16 used a \textit{k}-means clustering based on 39-dimensional MFCC features for the VDBO stems. Then for the mixture, for every 5 ms frame the system tried to identify which of the VDBO stems was dominant via the MFCC features, and then allocate the frame to the appropriate VDBO signal. However, such an approach struggles when more than one VDBO stem is prominent in a frame.

A statistical analysis of the ICASSP24 results was performed. First T16 was removed as an outlier as it's mean and standard deviation were both much smaller than all other systems. The systems using supplementary data or augmentation were also removed to ensure scores from each system were statistically independent (e.g. T31 was analysed but not T31-A). This left 13 systems.

As with CAD1, the data did not meet the normality assumption needed to use an ANOVA, and therefore an analysis of main effects using non-parametric approaches was used.

\begin{figure}
    \centering
    \includegraphics[width=\linewidth]{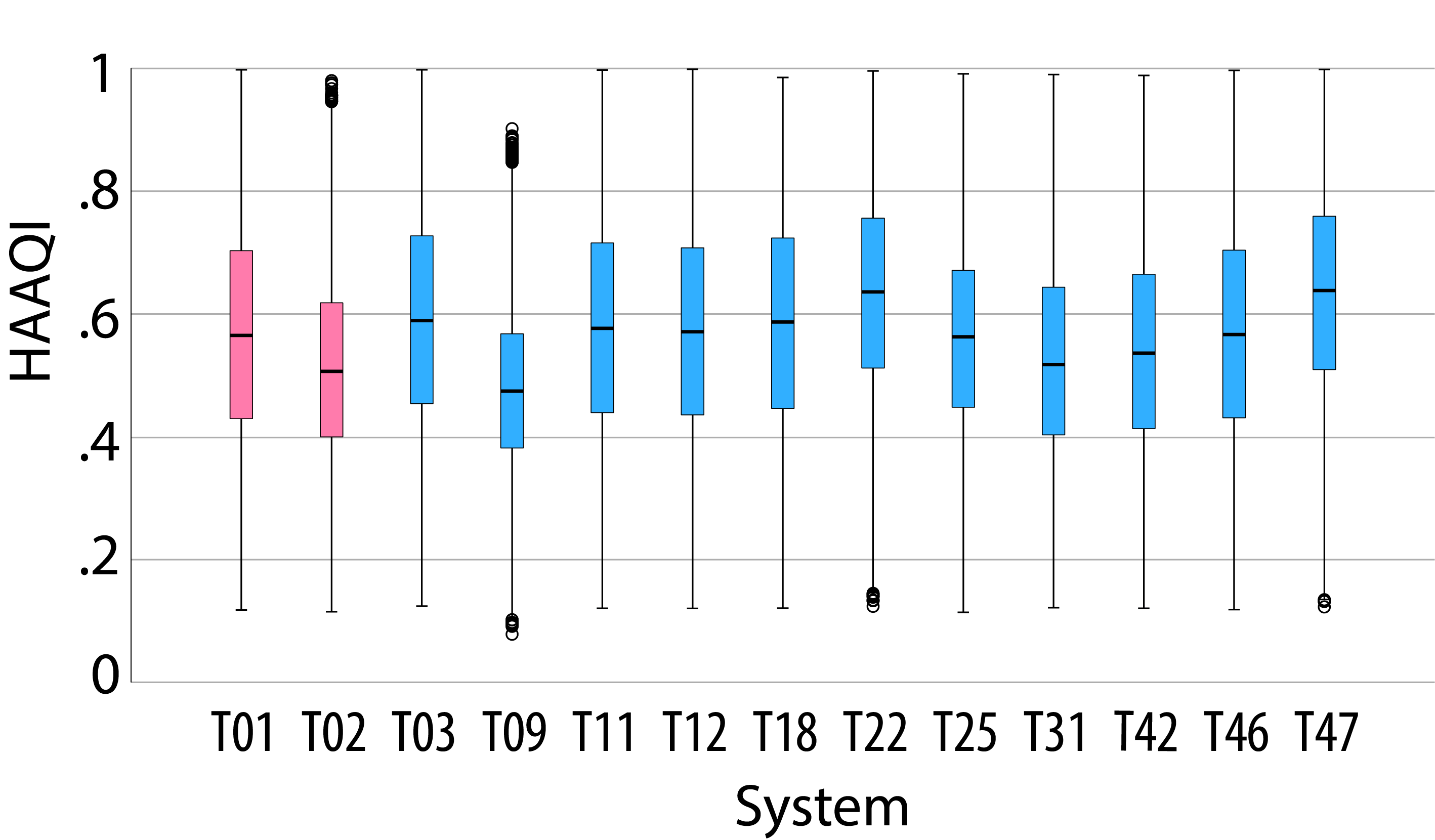}
    \caption{HAAQI scores vs system for ICASSP24. Baselines shown in pink.}
    \label{fig:systemboxploticassp24}
\end{figure}

Figure \ref{fig:systemboxploticassp24} shows a box-plot for the HAAQI scores vs system. A one-way Kruskal-Wallis test with HAAQI as the dependent variable and system as the independent variable was significant (N=249,600, df=12, H=13,682, p$<$0.001, 
$\eta ^2$=0.05). The effect size is small, however. Pairwise comparisons show that most systems were significantly different from each other. (p$<$0.001 for pairs with significant difference, except T25-T46 and T11-T18 where p = 0.03; Bonferroni correction for multiple tests was applied). The exceptions where there was no significant difference were: T01 and T12 (p=1); T01 and T046 (p=1); T01 and T25 (p=0.2); T11 and T12 (p=0.2); and T03 and T18 (p=0.2).

It was hypothesised that the greater the differences in gains applied to the VDBO stems before remixing, the poorer the performance would be. The thinking here was that any bleed or artefacts created during demixing will be more evident in the remix. To test this idea, the HAAQI scores were correlated with the standard deviation of the gains applied to the VDBO stems. The Spearman's rho was -0.318 (N=249,600; p$<$0.001). This indicates that HAAQI scores were indeed lower when there were larger differences in gains between the VDBO stems. This explained about 10\% of the rank variance.

The HAAQI scores were analysed to determine whether they varied with hearing loss severity. This analysis used the same hearing severity classifications as for CAD1. The Spearman's rho between HAAQI and hearing loss severity was -0.454 (N=249,600; p$<$0.001). This means that the more severe the hearing loss the lower the HAAQI scores. This explains 20\% of the rank variance. System T09 was the only one that did not have a linear relationship for hearing loss severity. For that system, the best scores were for moderate loss.

A greater hearing loss severity means that the NAL-R amplification would have been larger, especially at high frequencies. One possibility is that errors or artefacts in the demix are greater at higher frequencies, and hence the HAAQI score decreases for larger hearing loss severity. Another possible cause is the amplification creating more clipping in some music extracts with greater hearing loss severity, which would then decrease HAAQI.

Exploring how the HAAQI scores varied with the angle between the listener and the left and right loudspeakers in the stereo reproduction yielded a result that was significant but had tiny effect sizes. Significance is found because of the very large number of examples (N=249,600) but the differences in HAAQI scores were too small to be important.

\section{Discussion}

The Cadenza project has created the first series of machine-learning challenges to increase the number of music processing researchers considering hearing loss within their work. We have developed baseline software and curated databases released open-source without restriction. The aim is to catalyse a cultural shift in the audio machine learning community so more research includes the range of hearing abilities seen in the general population, rather than the default assumption of young 'normal hearing' \citep{drever2022aural}.

One difficulty with using the challenge methodology nowadays is the number of signal processing competitions being run make it harder to get a large number of entrants. The increase in the number of entrants from CAD1 to ICASSP24 shows that Cadenza is beginning to grow the community. This has been done by engaging with researchers who work on music demixing. Using gatekeepers to raise awareness of challenges is helpful, which is why we ran a challenge as part of ICASSP. Continuing this, the next Cadenza Challenge, CAD2, is an official challenge of the IEEE Signal Processing Society. It will also have some modest cash prizes to encourage entrants.

Choosing appropriate tasks, rules and evaluation methods for a new series is difficult because these are set before the challenge launches. The organisers must make an informed estimate of what teams might be able to achieve. In the CAD1 challenge, the baselines were based on state-of-the-art demixing models that, with the benefit of hindsight, did not leave enough room for improvement for challenge entrants. Learning from this, the ICASSP24 challenge made the problem more difficult by introducing loudspeaker reproduction and different gains for the VDBO stems in the remix. The lower scores for the top systems in ICASSP24 compared to the best in CAD1 indicate that there is still scope for further research into the ICASSP24 scenario.

Nearly all entrants used non-causal signal processing, which means the methods could only be applied to recorded music or broadcast situations where a delay in processing is not an issue. For hearing aids and live music, low latency and causal methods are required. In future, more work is needed to encourage causal systems. Hence, the second Cadenza challenge CAD2, features a causal baseline for entrants to beat.

The objective metrics currently available for machine learners to use during training also could be improved. HAAQI was used because it is the only audio quality metric that accounts for hearing loss and hearing aid processing. However, it is not ideal for machine learning because it is slow to compute and non-differentiable. Furthermore, as discussed above (see Materials - Evaluation) the amplification stage using in HAAQI sets the gold-standard for entrants to try and achieve, and there is no consensus on what that should be. Despite this, CAD2 will also move away from NAL-R to use non-linear amplification with parameter settings similar to those used in current hearing aids and the frequency-specific gains determined by each individual's pure-tone thresholds.

A non-intrusive metric developed from listening tests might overcome some of these issues. The audio created in CAD1 has been used in listening tests and work is ongoing to create a metric based on those results. For CAD2 one of the two tasks is improving lyric intelligibility.  For intelligibility assessment, we are using the Whisper model \citep{radford2022robust} to transcribe lyrics and compute word correct rates. Whisper does not require a reference signal.

In challenges, the databases that are available in the public domain also limit the tasks that can be set. This is especially true for challenges involving music because of copyright. CAD1/ICASSP24 was limited to mostly pop/rock music because of this. However, hearing loss is much more prevalent in older people, and our listening panel has a preference for classical music. For this reason, CAD2 will extend the demix/remix task to include classical music for small string and woodwind ensembles. This has required the synthesis of a new training dataset of classical music for  woodwind quartets \citep{cox2024cadenza}.

\section{Conclusions}

This paper details the first application of a machine learning challenge methodology to the problem of improving music for those with a hearing loss. The tasks focused on demixing and then remixing pop/rock music to allow the rebalancing of the instruments within the recording. We provided entrants with a common set of baseline tools, databases, evaluation metrics and challenge rules. While the design of the challenge built on previous demixing challenges, the addition of listeners with different hearing characteristics added complexity to the data, software baseline and evaluation metrics. A further innovation in the ICASSP24 challenge was the addition of loudspeaker listening and gains being applied to the separated stems before remixing. Loudspeaker reproduction made the separation of instruments more challenging due to frequency-dependent mixing of left and right signals and the gains applied to the tracks creates remix tracks that could highlight poorer separation. The machine learning methods used to demix the signals were nearly all refinements of current state-of-the art algorithms, either HDemucs or OpenUnmix.

The objective evaluation systems is always limited by the metric available, because this can only ever be an approximation to human listening. A true test of system requires listening tests. These have been carried out on the audio submitted to the CAD1 challenge. The design of these experiments and results will be presented in a companion paper.

The Cadenza challenge series was established to grow a community that includes hearing difference in their audio machine learning. It was pleasing to see that the number of systems entered roughly doubled between the two challenges. This has been achieved by tapping into the community of researchers already working on sound demixing. The next challenge, CAD2, includes a task on lyric intelligibility. The hope is that researchers who work on speech enhancement will be interested in adapting their algorithms to lyrics in music.

\section{Acknowledgements}
We thank our partners: BBC, Google, Logitech, RNID, Sonova, Universit\"{a}t Oldenburg.

\section{Declaration of conflicting interest}
The authors have no conflicting interest to declare.

\section{Funding statement}
The Cadenza project is supported by the Engineering and Physical Sciences Research Council (EPSRC) via grant number: EP/W019434/1.

\section{Ethical approval and informed consent statements}
The project has ethical approval from the University of Salford (no. 0718).

\section{Data availability statement}
The data from the challenges are available under a Creative Commons Attribution license \citep{roa_dabike_2024, roa_dabike_2024_ICASSP}.

\section{Copyright}

\bibliographystyle{SageH} 
\bibliography{bibliography.bib,bibliography_bid.bib}

\begin{thebibliography}{58}
\providecommand{\natexlab}[1]{#1}
\providecommand{\url}[1]{\texttt{#1}}
\providecommand{\urlprefix}{URL }
\expandafter\ifx\csname urlstyle\endcsname\relax
  \providecommand{\doi}[1]{DOI:\discretionary{}{}{}#1}\else
  \providecommand{\doi}{DOI:\discretionary{}{}{}\begingroup \urlstyle{rm}\Url}\fi

\bibitem[{Ahn et~al.(2021)Ahn, Choi, Kang, Choi, Lee, Lee, Hong and Moon}]{ahn2021influence}
Ahn J, Choi JE, Kang JY, Choi IJ, Lee MC, Lee BC, Hong SH and Moon IJ (2021) The influence of non-linear frequency compression on the perception of speech and music in patients with high frequency hearing loss.
\newblock \emph{Journal of audiology \& otology} 25(2): 80.

\bibitem[{Akeroyd et~al.(2023{\natexlab{a}})Akeroyd, Bailey, Barker, Cox, Culling, Graetzer, Naylor, Podwi{\'n}ska and Tu}]{akeroyd20232nd}
Akeroyd MA, Bailey W, Barker J, Cox TJ, Culling JF, Graetzer S, Naylor G, Podwi{\'n}ska Z and Tu Z (2023{\natexlab{a}}) The 2nd clarity enhancement challenge for hearing aid speech intelligibility enhancement: Overview and outcomes.
\newblock In: \emph{ICASSP 2023 IEEE International Conference on Acoustics, Speech and Signal Processing}. IEEE, pp. 1--5.

\bibitem[{Akeroyd et~al.(2023{\natexlab{b}})Akeroyd, Bailey, Barker, Cox, Culling, Graetzer, Naylor, Podwińska and Tu}]{CEC2}
Akeroyd MA, Bailey W, Barker J, Cox TJ, Culling JF, Graetzer S, Naylor G, Podwińska Z and Tu Z (2023{\natexlab{b}}) {The 2nd Clarity Enhancement Challenge for Hearing Aid Speech Intelligibility Enhancement: Overview and Outcomes}.
\newblock In: \emph{ICASSP 2023 - 2023 IEEE International Conference on Acoustics, Speech and Signal Processing (ICASSP)}. Rhodes Island, Greece: IEEE, pp. 1--5.
\newblock \doi{10.1109/ICASSP49357.2023.10094918}.

\bibitem[{Barker et~al.(2020)Barker, Akeroyd, Cox, Culling, Graetzer, Naylor and Porter}]{barker2017}
Barker JP, Akeroyd MA, Cox T, Culling J, Graetzer S, Naylor G and Porter E (2020) Open challenges for driving hearing device processing: lessons learnt from automatic speech recognition.
\newblock In: \emph{Speech in Noise Workshop 2020}.
\newblock \doi{10.1136/bmj.l1285}.

\bibitem[{Benjamin and Siedenburg(2023)}]{benjamin2023exploring}
Benjamin AJ and Siedenburg K (2023) Exploring level-and spectrum-based music mixing transforms for hearing-impaired listeners.
\newblock \emph{The Journal of the Acoustical Society of America} 154(2): 1048--1061.
\newblock \doi{10.1121/10.0020269}.

\bibitem[{Blacking(1995)}]{blacking1995music}
Blacking J (1995) \emph{Music, culture, and experience: Selected papers of John Blacking}.
\newblock University of Chicago Press.

\bibitem[{Bramsl{\o}w et~al.(2018)Bramsl{\o}w, Naithani, Hafez, Barker, Pontoppidan and Virtanen}]{bramslow2018improving}
Bramsl{\o}w L, Naithani G, Hafez A, Barker T, Pontoppidan NH and Virtanen T (2018) Improving competing voices segregation for hearing impaired listeners using a low-latency deep neural network algorithm.
\newblock \emph{The Journal of the Acoustical Society of America} 144(1): 172--185.

\bibitem[{Byrne and Dillon(1986)}]{byrne1986national}
Byrne D and Dillon H (1986) The {N}ational {A}coustic {L}aboratories'({NAL}) new procedure for selecting the gain and frequency response of a hearing aid.
\newblock \emph{Ear and hearing} 7(4): 257--265.

\bibitem[{Byrne et~al.(2001)Byrne, Dillon, Ching, Katsch and Keidser}]{byrne2001nal}
Byrne D, Dillon H, Ching T, Katsch R and Keidser G (2001) Nal-nl1 procedure for fitting nonlinear hearing aids: Characteristics and comparisons with other procedures.
\newblock \emph{Journal of the American academy of audiology} 12(01): 37--51.

\bibitem[{Chen et~al.(2024)Chen, Vekkot and Shukla}]{chen2024music}
Chen J, Vekkot S and Shukla P (2024) Music source separation based on a lightweight deep learning framework (dttnet: Dual-path tfc-tdf unet).
\newblock In: \emph{ICASP 2024 IEEE International Conference on Acoustics, Speech and Signal Processing}. IEEE, pp. 656--660.

\bibitem[{Cox and Roa~Dabike(2024)}]{cox2024cadenza}
Cox TJ and Roa~Dabike G (2024) Cadenza challenge (cad2): databases for rebalancing classical music task (1.0.0).
\newblock \doi{10.5281/zenodo.12664932}.

\bibitem[{Croghan et~al.(2014)Croghan, Arehart and Kates}]{croghan2014music}
Croghan NB, Arehart KH and Kates JM (2014) Music preferences with hearing aids: Effects of signal properties, compression settings, and listener characteristics.
\newblock \emph{Ear and hearing} 35(5): e170--e184.
\newblock \doi{10.1097/AUD.0000000000000056}.

\bibitem[{Daly(2024)}]{T022}
Daly M (2024) Remixing music for hearing aids using ensemble of fine-tuned source separators.
\newblock In: \emph{2024 IEEE International Conference on Acoustics, Speech, and Signal Processing Workshops (ICASSPW)}. pp. 109--110.
\newblock \doi{10.1109/ICASSPW62465.2024.10627557}.

\bibitem[{D{\'e}fossez(2021)}]{defossez2021hybrid}
D{\'e}fossez A (2021) Hybrid spectrogram and waveform source separation.
\newblock \emph{arXiv preprint arXiv:2111.03600} .

\bibitem[{Denk et~al.(2018)Denk, Ernst, Heeren, Ewert and Kollmeier}]{DenkHRTF}
Denk F, Ernst SM, Heeren J, Ewert SD and Kollmeier B (2018) {The Oldenburg Hearing Device (OlHeaD) HRTF Database}.
\newblock Technical report, University of Oldenburg.

\bibitem[{Drever and Hugill(2022)}]{drever2022aural}
Drever JL and Hugill A (2022) Aural diversity: General introduction.
\newblock In: \emph{Aural Diversity}. Routledge, pp. 1--12.

\bibitem[{Fabbro et~al.(2024)Fabbro, Uhlich, Lai, Choi, Mart{\'i}nez-Ram{\'i}rez, Liao, Gadelha, Ramos, Hsu, Rodrigues et~al.}]{fabbro2024sound}
Fabbro G, Uhlich S, Lai CH, Choi W, Mart{\'i}nez-Ram{\'i}rez M, Liao W, Gadelha I, Ramos G, Hsu E, Rodrigues H et~al. (2024) The sound demixing challenge 2023--music demixing track.
\newblock \emph{Transactions of the international Society for Music Information Retrieval} \doi{10.5334/tismir.171}.

\bibitem[{Gou et~al.(2021)Gou, Yu, Maybank and Tao}]{gou2021knowledge}
Gou J, Yu B, Maybank SJ and Tao D (2021) Knowledge distillation: A survey.
\newblock \emph{International Journal of Computer Vision} 129(6): 1789--1819.

\bibitem[{Grais et~al.(2021)Grais, Zhao and Plumbley}]{grais2021multi}
Grais EM, Zhao F and Plumbley MD (2021) Multi-band multi-resolution fully convolutional neural networks for singing voice separation.
\newblock In: \emph{2020 28th European Signal Processing Conference (EUSIPCO)}. IEEE, pp. 261--265.

\bibitem[{Greasley et~al.(2020)Greasley, Crook and Fulford}]{Greasley2020}
Greasley A, Crook H and Fulford R (2020) Music listening and hearing aids: perspectives from audiologists and their patients.
\newblock \emph{International Journal of Audiology} 59(9): 694--706.
\newblock \doi{10.1080/14992027.2020.1762126}.

\bibitem[{Hake et~al.(2023)Hake, B{\"u}rgel, Nguyen, Greasley, M{\"u}llensiefen and Siedenburg}]{hake2023development}
Hake R, B{\"u}rgel M, Nguyen NK, Greasley A, M{\"u}llensiefen D and Siedenburg K (2023) Development of an adaptive test of musical scene analysis abilities for normal-hearing and hearing-impaired listeners.
\newblock \emph{Behavior Research Methods} 15(11): 1--26.
\newblock \doi{10.3758/s13428-023-02279-y}.

\bibitem[{Han and Lee(2024)}]{T011}
Han C and Lee S (2024) Optimizing music source separation in complex audio environments through progressive self-knowledge distillation.
\newblock In: \emph{2024 IEEE International Conference on Acoustics, Speech, and Signal Processing Workshops (ICASSPW)}. pp. 13--14.
\newblock \doi{10.1109/ICASSPW62465.2024.10626965}.

\bibitem[{Hennequin et~al.(2020)Hennequin, Khlif, Voituret and Moussallam}]{hennequin2020spleeter}
Hennequin R, Khlif A, Voituret F and Moussallam M (2020) Spleeter: a fast and efficient music source separation tool with pre-trained models.
\newblock \emph{Journal of Open Source Software} 5(50): 2154.

\bibitem[{Holighaus et~al.(2012)Holighaus, D{\"o}rfler, Velasco and Grill}]{holighaus2012framework}
Holighaus N, D{\"o}rfler M, Velasco GA and Grill T (2012) A framework for invertible, real-time constant-q transforms.
\newblock \emph{IEEE Transactions on Audio, Speech, and Language Processing} 21(4): 775--785.

\bibitem[{Kates and Arehart(2015)}]{kates2015hearing}
Kates JM and Arehart KH (2015) The hearing-aid audio quality index (haaqi).
\newblock \emph{IEEE/ACM transactions on audio, speech, and language processing} 24(2): 354--365.
\newblock \doi{10.1109/TASLP.2015.2507858}.

\bibitem[{Kim et~al.(2021{\natexlab{a}})Kim, Ji, Yoon and Hwang}]{kim2021self}
Kim K, Ji B, Yoon D and Hwang S (2021{\natexlab{a}}) Self-knowledge distillation with progressive refinement of targets.
\newblock In: \emph{Proceedings of the IEEE/CVF international conference on computer vision 2021}. pp. 6567--6576.

\bibitem[{Kim et~al.(2021{\natexlab{b}})Kim, Choi, Chung, Lee and Jung}]{kim2021kuielab}
Kim M, Choi W, Chung J, Lee D and Jung S (2021{\natexlab{b}}) Kuielab-mdx-net: A two-stream neural network for music demixing.
\newblock \emph{arXiv preprint arXiv:2111.12203} .

\bibitem[{Lan et~al.(2024)Lan, Cheng, He, Chen and Du}]{T003}
Lan H, Cheng T, He M, Chen H and Du J (2024) The ustc system for cadenza 2024 challenge.
\newblock In: \emph{2024 IEEE International Conference on Acoustics, Speech, and Signal Processing Workshops (ICASSPW)}. pp. 57--58.
\newblock \doi{10.1109/ICASSPW62465.2024.10627147}.

\bibitem[{Liberman and Wayne(2020)}]{liberman2020human}
Liberman M and Wayne C (2020) Human language technology.
\newblock \emph{AI Magazine} 41(2): 22--35.
\newblock \doi{10.1609/aimag.v41i2.5297}.

\bibitem[{Liutkus et~al.(2017)Liutkus, St{\"o}ter, Rafii, Kitamura, Rivet, Ito, Ono and Fontecave}]{liutkus20172016}
Liutkus A, St{\"o}ter FR, Rafii Z, Kitamura D, Rivet B, Ito N, Ono N and Fontecave J (2017) The 2016 signal separation evaluation campaign.
\newblock In: \emph{Latent Variable Analysis and Signal Separation: 13th International Conference, LVA/ICA 2017, Grenoble, France, February 21-23, 2017, Proceedings 13}. Springer, pp. 323--332.

\bibitem[{Looi et~al.(2019)Looi, Rutledge and Prvan}]{looi2019music}
Looi V, Rutledge K and Prvan T (2019) Music appreciation of adult hearing aid users and the impact of different levels of hearing loss.
\newblock \emph{Ear and hearing} 40(3): 529--544.
\newblock \doi{10.1097/AUD.0000000000000632}.

\bibitem[{MacDonald et~al.(2013)MacDonald, Kreutz and Mitchell}]{macdonald2013music}
MacDonald R, Kreutz G and Mitchell L (2013) \emph{Music, health, and wellbeing}.
\newblock Oxford University Press.
\newblock \doi{10.1093/acprof:oso/9780199586974.001.0001}.

\bibitem[{Madsen and Moore(2014)}]{madsen2014music}
Madsen SM and Moore BC (2014) Music and hearing aids.
\newblock \emph{Trends in Hearing} 18: 2331216514558271.
\newblock \doi{10.1177/2331216514558271}.

\bibitem[{Madsen et~al.(2015)Madsen, Stone, McKinney, Fitz and Moore}]{madsen2015effects}
Madsen SM, Stone MA, McKinney MF, Fitz K and Moore BC (2015) Effects of wide dynamic-range compression on the perceived clarity of individual musical instruments.
\newblock \emph{The Journal of the Acoustical Society of America} 137(4): 1867--1876.
\newblock \doi{10.1121/1.4914988}.

\bibitem[{Mitsufuji et~al.(2022)Mitsufuji, Fabbro, Uhlich, St{\"o}ter, D{\'e}fossez, Kim, Choi, Yu and Cheuk}]{mitsufuji2022music}
Mitsufuji Y, Fabbro G, Uhlich S, St{\"o}ter FR, D{\'e}fossez A, Kim M, Choi W, Yu CY and Cheuk KW (2022) Music demixing challenge 2021.
\newblock \emph{Frontiers in Signal Processing} 1: 808395.
\newblock \doi{10.3389/frsip.2021.808395}.

\bibitem[{Moore(2016)}]{moore2016effects}
Moore BC (2016) Effects of sound-induced hearing loss and hearing aids on the perception of music.
\newblock \emph{Journal of the Audio Engineering Society} 64(3): 112--123.
\newblock \doi{10.17743/jaes.2015.0081}.

\bibitem[{Ono et~al.(2015)Ono, Rafii, Kitamura, Ito and Liutkus}]{Ono2015}
Ono N, Rafii Z, Kitamura D, Ito N and Liutkus A (2015) The 2015 signal separation evaluation campaign.
\newblock In: \emph{Proceedings of the International Conference on Latent Variable Analysis and Signal Separation (LVA/ICA)}. Liberec, Czech Republic, pp. 387--395.
\newblock \doi{10.1007/978-3-319-22482-4_45}.

\bibitem[{Park et~al.(2019)Park, Chan, Zhang, Chiu, Zoph, Cubuk and Le}]{park2019specaugment}
Park DS, Chan W, Zhang Y, Chiu CC, Zoph B, Cubuk ED and Le QV (2019) Specaugment: A simple data augmentation method for automatic speech recognition.
\newblock \emph{Interspeech} .

\bibitem[{Pereira et~al.(2023)Pereira, Araújo, Korzeniowski and Vogl}]{moisesdb}
Pereira I, Araújo F, Korzeniowski F and Vogl R (2023) Moisesdb: A dataset for source separation beyond 4-stems.
\newblock Preprint arXiv:2307.15913.

\bibitem[{Radford et~al.(2022)Radford, Kim, Xu, Brockman, McLeavey and Sutskever}]{radford2022robust}
Radford A, Kim JW, Xu T, Brockman G, McLeavey C and Sutskever I (2022) Robust speech recognition via large-scale weak supervision. arxiv (2022).
\newblock \emph{arXiv preprint arXiv:2212.04356} \doi{10.48550/arXiv.2212.04356}.

\bibitem[{Rafii et~al.(2019)Rafii, Liutkus, St{\"o}ter, Mimilakis and Bittner}]{musdb18-hq}
Rafii Z, Liutkus A, St{\"o}ter FR, Mimilakis SI and Bittner R (2019) {MUSDB18-HQ} - an uncompressed version of musdb18.
\newblock 10.5281/zenodo.3338373.

\bibitem[{Roa~Dabike et~al.(2024)Roa~Dabike, Akeroyd, Bannister, Barker, Cox, Fazenda, Firth, Graetzer, Greasley, Vos and Whitmer}]{ICASSP2024}
Roa~Dabike G, Akeroyd MA, Bannister S, Barker J, Cox TJ, Fazenda B, Firth J, Graetzer S, Greasley A, Vos RR and Whitmer WM (2024) The icassp sp cadenza challenge: Music demixing/remixing for hearing aids.
\newblock In: \emph{2024 IEEE International Conference on Acoustics, Speech, and Signal Processing Workshops (ICASSPW)}. pp. 93--94.
\newblock \doi{10.1109/ICASSPW62465.2024.10626340}.

\bibitem[{Roa~Dabike et~al.(2023)Roa~Dabike, Bannister, Firth, Graetzer, Vos, Akeroyd, Barker, Cox, Fazenda, Greasley and Whitmer}]{CAD1}
Roa~Dabike G, Bannister S, Firth J, Graetzer S, Vos RR, Akeroyd MA, Barker J, Cox TJ, Fazenda B, Greasley A and Whitmer WM (2023) {The First Cadenza Signal Processing Challenge: Improving Music for Those With a Hearing Loss}.
\newblock In: \emph{2nd Workshop on Human-Centric Music Information Research (HCMIR) 2023}.

\bibitem[{Roa~Dabike and Cox(2024{\natexlab{a}})}]{roa_dabike_2024}
Roa~Dabike G and Cox TJ (2024{\natexlab{a}}) Cadenza challenge (cad1): databases for the first cadenza challenge - task1.
\newblock \doi{10.5281/zenodo.13285384}.

\bibitem[{Roa~Dabike and Cox(2024{\natexlab{b}})}]{roa_dabike_2024_ICASSP}
Roa~Dabike G and Cox TJ (2024{\natexlab{b}}) Cadenza challenge (icassp24): databases for icassp 2024 cadenza grand challenge (1.0.0).
\newblock \doi{10.5281/zenodo.13285307}.

\bibitem[{Rouard et~al.(2023)Rouard, Massa and D{\'e}fossez}]{rouard2023hybrid}
Rouard S, Massa F and D{\'e}fossez A (2023) Hybrid transformers for music source separation.
\newblock In: \emph{ICASSP 2023 IEEE International Conference on Acoustics, Speech and Signal Processing}. IEEE, pp. 1--5.

\bibitem[{Sawata et~al.(2021)Sawata, Uhlich, Takahashi and Mitsufuji}]{sawata2021all}
Sawata R, Uhlich S, Takahashi S and Mitsufuji Y (2021) All for one and one for all: Improving music separation by bridging networks.
\newblock In: \emph{IEEE 2021 International Conference on Acoustics, Speech and Signal Processing (ICASSP)}. IEEE, pp. 51--55.

\bibitem[{Shao et~al.(2024)Shao, Chen and Dubnov}]{T046}
Shao K, Chen K and Dubnov S (2024) Music enhancement with deep filters: A technical report for the icassp 2024 cadenza challenge.
\newblock In: \emph{2024 IEEE International Conference on Acoustics, Speech, and Signal Processing Workshops (ICASSPW)}. pp. 119--120.
\newblock \doi{10.1109/ICASSPW62465.2024.10626759}.

\bibitem[{Siedenburg et~al.(2020)Siedenburg, R{\"o}ttges, Wagener and Hohmann}]{siedenburg2020can}
Siedenburg K, R{\"o}ttges S, Wagener KC and Hohmann V (2020) Can you hear out the melody? testing musical scene perception in young normal-hearing and older hearing-impaired listeners.
\newblock \emph{Trends in Hearing} 24: 2331216520945826.
\newblock \doi{10.1177/2331216520945826}.

\bibitem[{Stevens et~al.(2013)Stevens, Flaxman, Brunskill, Mascarenhas, Mathers and Finucane}]{stevens2013global}
Stevens G, Flaxman S, Brunskill E, Mascarenhas M, Mathers CD and Finucane M (2013) Global and regional hearing impairment prevalence: an analysis of 42 studies in 29 countries.
\newblock \emph{The European Journal of Public Health} 23(1): 146--152.

\bibitem[{St{\"o}ter et~al.(2018)St{\"o}ter, Liutkus and Ito}]{stoter20182018}
St{\"o}ter FR, Liutkus A and Ito N (2018) The 2018 signal separation evaluation campaign.
\newblock In: \emph{Latent Variable Analysis and Signal Separation}. Springer, pp. 293--305.

\bibitem[{St{\"o}ter et~al.(2019)St{\"o}ter, Uhlich, Liutkus and Mitsufuji}]{stoter19}
St{\"o}ter FR, Uhlich S, Liutkus A and Mitsufuji Y (2019) {Open-Unmix} - a reference implementation for music source separation.
\newblock \emph{J. Open Source Softw.} \doi{10.21105/joss.01667}.

\bibitem[{Uys et~al.(2012)Uys, Pottas, Vinck and Van~Dijk}]{uys2012influence}
Uys M, Pottas L, Vinck B and Van~Dijk C (2012) The influence of non-linear frequency compression on the perception of music by adults with a moderate to severe hearing loss: Subjective impressions.
\newblock \emph{South African Journal of Communication Disorders} 59(1): 53--67.
\newblock \doi{10.4102/sajcd.v59i1.22}.

\bibitem[{Vaisberg et~al.(2019)Vaisberg, Martindale, Folkeard and Benedict}]{vaisberg2019qualitative}
Vaisberg JM, Martindale AT, Folkeard P and Benedict C (2019) A qualitative study of the effects of hearing loss and hearing aid use on music perception in performing musicians.
\newblock \emph{Journal of the American Academy of Audiology} 30(10): 856--870.
\newblock \doi{10.3766/jaaa.17019}.

\bibitem[{Vatakis and Spence(2006)}]{vatakis2006audiovisual}
Vatakis A and Spence C (2006) Audiovisual synchrony perception for music, speech, and object actions.
\newblock \emph{Brain research} 1111(1): 134--142.

\bibitem[{von Gablenz et~al.(2017)von Gablenz, Hoffmann and Holube}]{Von2017}
von Gablenz P, Hoffmann E and Holube I (2017) Prevalence of hearing loss in {Northern and Southern Germany}.
\newblock \emph{{HNO}} 65(Suppl 2): 130--135.
\newblock \doi{10.1007/s00106-016-0318-4}.

\bibitem[{{World Health Organization}(2021)}]{world2021world}
{World Health Organization} (2021) \emph{World report on hearing}.
\newblock World Health Organization.

\bibitem[{Yin et~al.(2024)Yin, Wang, Bai, Shi, Gan and Chen}]{T018}
Yin H, Wang M, Bai J, Shi D, Gan WS and Chen J (2024) Sub-band and full-band interactive u-net with dprnn for demixing cross-talk stereo music.
\newblock In: \emph{2024 IEEE International Conference on Acoustics, Speech, and Signal Processing Workshops (ICASSPW)}. pp. 21--22.
\newblock \doi{10.1109/ICASSPW62465.2024.10627597}.

\end{thebibliography}

\end{document}